\documentstyle[12pt]{article}

\newcommand{\be}{\begin{equation}}
\newcommand{\ee}{\end{equation}}
\newcommand{\bea}{\begin{eqnarray}}
\newcommand{\eea}{\end{eqnarray}}

\def\sig{\sigma}

\newcommand\half{\frac{1}{2}}

\begin{document}
\setlength{\baselineskip}{0.7cm}

\begin{titlepage}
\null
\begin{flushright}
KEK-TH-931\\
hep-th/0312148 \\
December, 2003
\end{flushright}
\vskip 1cm
\begin{center}
{\LARGE\bf 
Supersymmetric Radius Stabilization\\ 

\vspace{3mm}

in Warped Extra Dimensions
} 

\lineskip .75em
\vskip 1.5cm

\normalsize

{\large\bf Nobuhito Maru$^{a}$
\footnote{Special Postdoctoral Researcher} 
\footnote{E-Mail: maru@postman.riken.go.jp} } 
 and {\large\bf Nobuchika Okada$^{b}$ 
\footnote{E-Mail: okadan@post.kek.jp}
}

\vspace{10mm}

{\it $^{a}$Theoretical Physics Laboratory, RIKEN, \\
 2-1 Hirosawa, Wako, Saitama 351-0198, JAPAN} \\
\vspace*{0.5cm}
{\it $^{b}$Theory Group, KEK, \\
1-1 Oho, Tsukuba, Ibaraki 305-0801, JAPAN} \\

\vspace{18mm}

{\bf Abstract}\\[5mm]
{\parbox{13cm}{\hspace{5mm}
%
We propose a simple model of extra-dimensional radius 
stabilization in a supersymmetric Randall-Sundrum model. 
In our model, we introduce only a bulk hypermultiplet 
and source terms (tadpole terms) on each boundary branes. 
With appropriate choice of model parameters, 
we find that the radius can be stabilized 
by supersymmetric vacuum conditions. 
Since the radion mass can be much larger than 
the gravitino mass and even the original 
supersymmetry breaking scale, radius stability is ensured 
even in the presence of supersymmetry breaking. 
We find a parameter region in which unwanted scalar masses 
induced by quantum corrections through the bulk hypermultiplet 
and a bulk gravity multiplet 
are suppressed and the anomaly mediation contribution dominates. 

}}

\end{center}

\end{titlepage}

\section{Introduction}
Motivated by an alternative solution to the hierarchy problem, 
much attention has been recently paid to the brane world scenario 
\cite{ADD, RS1, RS2}. 
In this scenario, the hierarchy between the weak scale and 
the Planck scale is geometrically obtained by the presence of 
large extra spatial dimensions \cite{ADD} or warped extra spatial dimensions 
\cite{RS1,RS2} without supersymmetry (SUSY).

There is another motivation to consider the brane world scenario 
in the context of SUSY breaking mediation in supergravity (SUGRA) 
as first discussed in \cite{RS}. 
In 4D SUGRA, once SUSY is broken in the hidden sector, 
SUSY breaking effects can be mediated to the visible sector automatically 
through the Planck suppressed SUGRA contact interactions, 
\bea
\label{4Dscalar}
\int d^4 \theta c_{ij} \frac{Z^\dag Z Q^\dag_i Q_j}{M_4^2} 
\to c_{ij}m_{3/2}^2 \tilde{Q}^\dag_i \tilde{Q}_j, 
\eea
we obtain soft scalar masses of the order of 
the gravitino mass $m_{3/2}$ for the scalar partners. 
Here $Z$ is a SUSY breaking chiral superfield with $F_Z \ne 0$, 
$Q_i$ is the minimal SUSY standard model (MSSM) chiral superfields 
of $i$-th flavor, 
$\tilde{Q}_i$ is its scalar component, 
$c_{ij}$ are flavor dependent constants 
and $M_4$ is a 4D Planck scale. 
Although the soft SUSY breaking masses 
are severely constrained to be almost flavor diagonal by experiments, 
there is no symmetry reason for $c_{ij}=\delta_{ij}$ in 4D SUGRA. 
Therefore 4D SUGRA model suffers from the so-called SUSY FCNC problem. 
Recently, it was proposed that 
the direct contact terms such as (\ref{4Dscalar}) 
between the visible and the hidden sectors are naturally suppressed 
if the two sectors are separated each other 
along the direction of extra spatial dimensions \cite{RS,LS1}. 
This is because the higher dimensional locality forbids 
the direct contact term. 
This scenario is called the ``sequestering scenario''. 
In this setup, soft SUSY breaking terms in the visible sector 
are generated through a superconformal anomaly (anomaly mediation)  
and the resultant mass spectrum is found to be flavor-blind, 
there is no SUSY FCNC problem \cite{RS,GLMR}.\footnote{
If the visible sector is the MSSM, the sleptons are found to be tachyonic. 
There are many proposals for non-minimal models providing 
a realistic mass spectrum \cite{negativeslepton}.} 
Thus it is well motivated to consider the SUSY brane world scenario.

In the brane world scenario, there is an important issue 
called ``radius stabilization''. 
In order for the scenario to be phenomenologically viable, 
the compactification radius should be stabilized. 
However in the normal SUSY brane world scenario, 
the ``radion'', a scalar field parameterizing the compactification radius, 
is found to be a moduli field if SUSY is manifest, 
and the radius is undetermined. 
Although the nontrivial radion potential emerges once SUSY is broken, 
such a potential usually destabilize the radius. 
While some fields introduced in the bulk may work 
to stabilize the radius, 
these new fields might generate new flavor violating 
soft SUSY breaking terms in the visible sector 
larger than the anomaly mediation contributions. 
For the above discussions, see \cite{LO} and \cite{PP} for example. 
Unfortunately, this situation seems to be generic 
in the SUSY brane world scenario. 
Therefore, when we construct a realistic SUSY brane world model, 
we have to consider SUSY breaking, its mediation mechanism 
and the radius stabilization all together from the beginning. 
This makes a model construction very hard. 
We need a simple model which can stabilize the radius 
independently of the SUSY breaking and its mediation mechanism. 

In this paper, we propose a simple model of 
extra-dimensional radius stabilization 
in a SUSY Randall-Sundrum model. 
We introduce only a bulk hypermultiplet and source terms 
(tadpole terms) on each boundary branes. 
With appropriate values of the source terms 
and a mass of bulk hypermultiplet, 
we can find a classical SUSY configuration 
connecting two branes, and the radius is 
completely determined by a SUSY vacuum condition. 
The radion mass can be much larger 
than the gravitino mass and 
even the original SUSY breaking scale. 
We will show that the radion potential does not 
receive SUSY breaking effects so much, 
and the radius stability is ensured even with SUSY breaking. 
Unwanted soft scalar masses induced by 
quantum corrections through the bulk hypermultiplet 
and the bulk gravity multiplet are estimated. 
We find a parameter region in which they are suppressed 
and the anomaly mediation contribution dominates. 
Based on our model, we can discuss 
the radius stabilization problem 
independently of SUSY breaking and its mediation mechanism,
and the original picture of the sequestering scenario can work. 
This is a remarkable advantage for model buildings 
in SUSY brane world scenario.

For the related works, see \cite{GW, AHSW, LS1, LS2, PP, GLN} for example. 
We give some brief comments on relations between our model and 
models of Goldberger and Wise \cite{GW}, Arkani-Hamed et al. \cite{AHSW} 
and Goh, Luty and Ng \cite{GLN}. 
Our model may be understood as a SUSY version of \cite{GW} in some sense. 
In both models, a classical configuration of the bulk scalar field 
(hypermultiplet in our case) connecting 
two boundary branes stabilizes the radius 
by adjusting parameters on the boundaries and in the bulk. 
The radius stability is ensured by SUSY in our model. 
As will be seen, our model is similar to the model in Ref.~\cite{AHSW}. 
While in \cite{AHSW} the radius stabilization is discussed 
in the global SUSY theory with the massive hypermultiplet in the bulk 
in flat space-time background, 
our model is based on 5D SUGRA with the Randall-Sundrum background. 
Even if taking a flat limit, our model does not reduce 
into the model in Ref.~\cite{AHSW}, since the hypermultiplet 
in our model becomes massless in this limit. 
Our model is also similar to the model in Ref~\cite{GLN}. 
The radius stabilization is realized with SUSY breaking in their model, 
while in our model it is realized in a supersymmetric way.

This paper is organized as follows. 
In the next section, 
we introduce our model and discuss how the radius is stabilized. 
Then, the radion mass is calculated in section 3 
and it turns out to be very heavy. 
In section 4, it is shown that the radius stability is ensured 
even if we take SUSY breaking effects into account. 
In section 5, we estimate unwanted scalar masses induced 
by quantum corrections through the bulk hypermultiplet 
and the bulk gravity multiplet. 
We find a parameter region in order for our model 
to be phenomenologically viable. 
Section 6 is devoted to summary.

\section{Simple model of radius stabilization}

The starting point of our discussion is the following Lagrangian\footnote{
This Lagrangian is the one originated from 
the linearized supergravity (see for example Ref. \cite{LLP}). 
Considering that nonlinear terms 
in a full five dimensional supergravity are suppressed 
by the Planck scale $M_5$ and, as will be seen later, 
we can take the parameters in our model 
being much smaller than the Planck scale $M_5$, 
we can expect their effects negligible. 
Therefore, the Lagrangian is a good starting point 
of our arguments.}in five dimensional Randall-Sundrum background \cite{RS1,RS2}, 
in which the fifth dimension is compactified on an orbifold $S^1/Z_2$, 
\bea 
\label{originalL}
{\cal L}_5 &=& \int d^4 \theta \frac{T+T^\dagger}{2} 
e^{-(T+T^\dagger)\sig}
\left[ -6 M_5^3 + |H|^2 + |H^c|^2 \right] |\phi|^2 \nonumber \\
&+& \left[ \int d^2 \theta \phi^3 e^{-3T \sig} 
H \left\{
 \left( -\partial_y + \left( \frac{3}{2} + c \right) 
T \sig' \right) H^c + W_b(y) \right\} + {\rm h.c.}
\right] 
\eea
where five dimensional spacetime metric is given by 
\be
ds^2 = e^{-2 r \sig(y)} \eta_{\mu\nu}dx^\mu dx^\nu 
- r^2dy^2, ~(\mu, \nu=0,1,2,3)
\ee
where $r$ is the radius of the fifth dimension, 
 $0 \leq y \leq \pi$ is the angle on $S^1$, 
 and  $\sig(y) = k|y|$ with $k$ being an $AdS_5$ curvature scale. 
The prime denotes the differentiation 
with respect to $y$, $T$ is a radion chiral multiplet 
whose real part of scalar component gives the radius $r$, 
$\phi=1+\theta^2 F_\phi$ is a compensating multiplet, 
$H$ and $H^c$ are hypermultiplet components 
in terms of superfield notation in N=1 SUSY 
in four dimensions \cite{AGW,MP}, 
and $Z_2$ parity for $H$ and $H^c$ are defined 
as even and odd, respectively. 
$W_b \equiv J_0 \delta(y) - J_\pi \delta(y-\pi)$, 
where $J_{0,\pi}$ are constant source terms 
on each boundary branes at $y = 0, \pi$. 
Rescaling
\bea
(H, H^c) \to \frac{1}{\omega} (H, H^c), 
\quad \omega \equiv \phi e^{-T \sig}, 
\eea
we obtain more convenient form such as 
\bea
\label{rescaled}
{\cal L}_5 &\to&
\int d^4 \theta \left[ -3 M_5^3 (T+T^\dagger) |\omega|^2 
+ \frac{T+T^\dagger}{2}(|H|^2 + |H^c|^2) \right] \nonumber \\
&&+ \left[ \int d^2 \theta \omega H \left\{ - \partial_y H^c
+ \left( c + \half \right) 
T \sig' H^c + \omega W_b \right\} + {\rm h.c.} \right]. 
\label{rescaled2}
\eea

Supersymmetric configurations are easily obtained 
from F-flatness conditions, 
\footnote{It is well-known that SUSY vacua in global SUSY theory 
are also SUSY vacua in supergravity if the VEV of the superpotential 
vanishes at the minimum \cite{Weinberg}. This fact can be shown 
in an elegant way by use of superconformal framework. }  
\bea
0 &=&  - \partial_y H^c + \left( c + \half \right) 
T \sig' H^c + e^{-T\sig} W_b, \\ 
0 &=& \partial_y H + \left( c-\frac{1}{2} \right) T \sig' H. 
\eea
It is useful to parameterize 
 the $Z_2$ odd field as 
 $ H^c(y)= \varepsilon(y) \tilde{H}^c(y)$ 
 with a step function $\varepsilon(y)=-1,+1$ for $ y < 0,  0< y$ 
 and a regular function $\tilde{H}^c(y)$. 
Except the boundary points $y=0, \pi$, 
the solutions can be easily found as  
\bea
\label{SUSYcfg}
H(y) = C_H e^{(\half-c)T\sig},\quad 
\tilde{H}^c(y) =C_{\tilde{H}^c} e^{(c+\half)T\sig},  
\eea
with integration constants, $C_H$ and $C_{\tilde{H}^c}$.  
The source terms on each boundaries lead 
to the boundary conditions for $\tilde{H}^c$ such as 
\be
\tilde{H}^c(0) = \frac{J_0}{2}, \quad 
\tilde{H}^c(\pi) = \frac{J_\pi}{2}e^{-Tk\pi}.   
\ee
As a result, we obtain the SUSY vacuum condition of the form 
\be
\label{SUSYcondition}
 J_0 -  J_\pi e^{-\left(\frac{3}{2}+c\right)Tk\pi}= 0. 
\ee
Thus, the radius is determined 
with appropriate values of $J_{0,\pi}$ and 
the bulk hypermultiplet mass $c$. 
This is the point of this paper.

\section{4D effective action and radion Mass}
It is convenient to describe our model in the form of 
4D effective theory with only the light hypermultiplet. 
Substituting the light mode wave functions 
for the hypermultiplet,    
\bea
H(x,y) &=& h(x) e^{(\half-c)T \sig}, \\
H^c(x,y) &=& h^c(x) \varepsilon(y) e^{(c+\half)T \sig},
\eea
into (\ref{rescaled2}) and performing $y$ integration, 
we obtain the effective K\"ahler potential part 
\bea
\label{effK}
\int d^4 \theta {\cal K}_{\rm{eff}} = 
\int d^4 \theta \left[ 
f(T,T^\dag)|\phi|^2 + K(T,T^\dag)|h|^2 + K^c(T,T^\dag)|h^c|^2 
\right]
\eea
where 
\bea 
f(T,T^\dag) &=&  -\frac{3M_5^3}{k} 
\left[ 1-e^{-(T+T^\dag)k \pi} \right], \nonumber \\ 
K(T,T^\dag) &=& \frac{e^{(\half-c)(T+T^\dag)k \pi}-1}{(1-2c)k},
 \nonumber \\
K^c(T,T^\dag)&=& 
 \frac{e^{(\half+c)(T+T^\dag)k \pi}-1}{(1+2c)k}, 
\eea
and the effective superpotential part 
\bea
\label{effW}
\int d^2 \theta \phi^2 W(h,T)=
\int d^2 \theta 
\phi^2 h \left[ J_0 - J_\pi e^{-(c+\frac{3}{2})Tk \pi} \right]. 
\eea
The SUSY vacuum condition $\partial W /\partial h =0$ 
leads to the same condition as (\ref{SUSYcondition}) 
as it should be. 
It is somewhat complicated but straightforward 
to calculate the scalar potential and found that 
the potential minimum at $T=T_0$ satisfying (\ref{SUSYcondition}), 
$h=0$ and arbitrary $h^c$. 
Since the effective superpotential is independent of $h^c$, 
$h^c$ is left undetermined. 
At the point $h=0$, the radion potential is found to be 
\bea
\label{SUSYpot}
V_{{\rm radion}} 
=  K(T,T^\dag)^{-1} 
\left| \frac{\partial W(h,T)}{\partial h} \right|^2 
=\frac{(1-2c) k }{e^{(\half-c)(T+T^\dag)k \pi}-1} 
\left| J_0 - J_\pi e^{-(c+\frac{3}{2}) Tk \pi} \right|^2 .
\eea
One can explicitly see that the potential minimum is 
given by the SUSY condition (\ref{SUSYcondition}). 
Note that $T \to \infty$ also gives the potential minimum. 
However this originates from the singularity of K\"ahler potential
$K(T,T^\dag)$, and thus this vacuum is not well-defined. 
It is interesting to take the flat limit, $k \to 0$. 
The scalar potential reduces to a runaway potential 
$V_4 \sim 1/(T+T^\dag)$, namely, the radius is not stabilized. 
This means that the warped background metric is crucial 
for the radius stabilization. 

Now we calculate a radion mass. 
Considering canonical normalization of the radion kinetic term, 
we can estimate the radion mass such as  
\bea
\label{radionmass1}
m^2_{{\rm radion}} &\sim & 
\left( \frac{\partial^2 f(T,T^\dag)}
{\partial T^\dag \partial T} \right)^{-1}
\left.  
\frac{\partial^2 V_{{\rm radion}} }{\partial T^\dag \partial T} 
 \right|_{T=T_0} \nonumber \\ 
&=& 
\frac{(1-2c)}{e^{(\half-c)(T+T^\dag)k \pi}-1} 
\left(\frac{(\frac{3}{2}+c)^2 |J_\pi|^2}{3 M_5^3} \right) 
k^2 
\left. 
e^{-(\half+c) (T+T^\dag) k \pi } \right|_{T=T_0} > 0
\eea
Note that the radion mass squared is always positive 
irrespective of the value of $c$. 
This means that the radius is stabilized and 
the configuration under consideration is stable. 
As an example, if we take\footnote{
In our model, the gauge hierarchy problem is solved by SUSY, 
so that it does not need to take $e^{-T k \pi} \simeq 10^{-16}$ 
as in the original Randall-Sundrum model.} 
$c = \half $, $e^{-T_0 k \pi} \sim 10^{-2}, 
J_\pi \sim (0.1 \times M_5)^{3/2}$ and $k \sim 0.1 \times M_5$, 
we obtain the radion mass,  
\bea
\label{radionmass3}
m_{{\rm radion}}^2 \sim (10^{-5} \times M_4)^2 
 \gg m_{3/2}^2, F_{{\rm hidden}}, 
\eea
which is much larger than the gravitino mass ($\sim 10$ TeV) 
in anomaly mediation scenario 
and the original SUSY breaking F-term scale 
$F_{{\rm hidden}}\sim m_{3/2} M_4 $ in a hidden sector. 
This fact implies that SUSY breaking effects little affect 
the radion potential, and the radius is not destabilized 
even in the presence of SUSY breaking. 
In the next section, we will check this expectation in more detail.

\section{Stability of radius under SUSY breaking effects}

Suppose that the hidden sector fields and visible sector fields 
reside on the each branes at the boundaries $y=0$ and $y=\pi$, 
respectively. 
In this setup, the hidden sector fields couple 
only to the compensating multiplet, and the other fields 
can be regarded as the visible sector fields. 
Once SUSY is broken in the hidden sector, 
the SUSY breaking effects emerge in the visible sector 
only through the non-vanishing $F_\phi$, 
and we can treat the compensating multiplet as a spurion. 
In order to prove the stability of the radius 
in presence of the SUSY breaking effects, 
we have to solve equations of motion for $H, H^c$ 
with non-vanishing $F_\phi$. 
However, it is hard to solve these complicated equations. 
Instead of solving them, we prove the radius stability 
in the effective 4D theory as an approximation 
since the effect of the small $F_\phi$ is important 
only for light fields. 
 
With the compensating multiplet $\phi=1+\theta^2 F_\phi$ 
as the spurion, 
the Lagrangian for the auxiliary fields can be read off 
from (\ref{effK}) and (\ref{effW}) such as
\bea
{\cal L}_{{\rm aux}} &=& 
F_T^\dag \left[ 
 (f_{T T^\dag} + K^c_{TT^\dag} |h^c|^2 + K_{TT^\dag} |h|^2) F_T 
 + (K^c_T h^c)^\dag F^c + (K_T h)^\dag F  
 + W_T^\dag + f_{T^\dag} F_\phi \right] \nonumber \\
&+& 
F^{c\dag} 
 \left[ (K^c_T h^c)  F_T + K^c F^c  \right] 
+ F^\dag \left[ (K_T h) F_T + K F + W_h^\dag \right]  \nonumber \\
&+& 
 F W_h + F_T W_T +2 ( F_\phi W + {\rm{h.c.}}) 
    + F_\phi^\dag f_T F_T  +  |F_\phi|^2 f  , 
\eea
where $ f_T $ stands for $\partial f/ \partial T$ etc. 
In the following, we estimate the deviation from the SUSY case 
in the first order of $F_\phi$. 
In this approximation, all the F-terms and the shifts 
of the field VEVs, $\delta h$ and $\delta T = T -T_0$, 
around the SUSY vacuum 
are regarded as ${\cal O}(F_\phi)$ variables. 
Thus we can approximate equations of motion for auxiliary fields as 
\bea
0&=&
(f_{T T^\dag} + K^c_{TT^\dag} |h^c|^2 + K_{TT^\dag} |h|^2) F_T 
+ (K^c_T h^c)^\dag F^c + (K_T h)^\dag F  
+ W_T^\dag + f_{T^\dag} F_\phi  \nonumber \\
&\sim &  
 (f_{T T^\dag} + K^c_{TT^\dag} |h^c|^2) F_T 
 + (K^c_T h^c)^\dag F^c + \delta h^\dag W_{hT}^\dag 
 + f_{T^\dag} F_\phi \\
0&=& (K^c_T h^c)  F_T + K^c F^c  \\
0 &=& (K_T \delta h) F_T + K F + W_h^\dag 
\sim  K F + W_{h T}^\dag \delta T^\dag . 
\eea
The solution $F_T$ and $F$ are given by 
\bea
F_T &\sim & -\frac{1}{C_T} \left.  
\left(\delta h^\dag  W_{h T}^\dag + f_{T^\dag} F_\phi 
\right) \right|_{T=T_0, h=0}  ,  \\ 
F &\sim & -\frac{1}{K} 
\left. W_{h T}^\dag \delta T^\dag   \right|_{T=T_0, h=0} ,
\eea
where 
\bea
C_T = f_{TT^\dag} + 
\left( K^c_{TT^\dag} -\frac{|K^c_T|^2}{K^c} \right) |h^c|^2  . 
\eea
Up to the second order, the scalar potential is given by 
\bea
\Delta V &=& - F W_h - F_T W_T - 2 ( F_\phi W + {\rm{h.c.}}) 
      - F_\phi^\dag f_T F_T  -  |F_\phi|^2 f  \nonumber \\ 
&\sim & 
\frac{1}{K}  |W_{h T} \delta T|^2 
+\frac{1}{C_T} | \delta h^\dag W_T^\dag - f_{T^\dag} F_\phi|^2  
 -|F_\phi|^2 f , 
\eea
and minimization conditions, 
$\frac{\partial \Delta V}{\partial \delta T} =0$ and 
$\frac{\partial \Delta V}{\partial \delta h} =0$, 
lead to  
\bea
&&\delta T \sim 0, \\
&&\delta h \sim - \frac{f_T(T_0)}{W_{h T}(T_0)} F_\phi^\dag 
\sim 
\frac{6}{2 c+3} 
\frac{M_5^3 F_\phi^\dag}{J_0 k} 
e^{-(T_0+T_0^\dag) k \pi} . 
\eea
Therefore, with appropriate values of parameters, 
the deviations are small enough for our treatment 
to be consistent. 
$h^c$ still remains undetermined. 
Numerical calculations shows 
that the above results gives good approximations. 
Now we have proven the radius stability 
even with the presence of SUSY breaking.

\section{Scalar masses induced by bulk fields}

We have introduced the hypermultiplet in the bulk 
for the radius stabilization. 
In general, there is a possibility that 
the flavor dependent soft SUSY breaking terms 
being phenomenologically dangerous  
are induced through the bulk hypermultiplet, 
since the ($Z_2$ even) hypermultiplet 
can directly couple to both the hidden and visible sector fields. 

Let us consider the effective K\"ahler potentials 
on the hidden brane at $y=0$ and 
the visible brane at $y=\pi$ such that 
(in the original basis of (\ref{originalL})) 
\bea
\label{hkahler}
 {\cal L}_{{\rm hidden}} 
 &=& \int_{0}^{\pi} dy\int d^4 \theta e^{-(T_0+T_0^\dag) \sigma} 
 \left[Z^\dag Z + 
 \frac{H_0^\dag H_0 Z^\dag Z }{M_5^3} \right] \delta(y) , \\
\label{vkahler} 
 {\cal L}_{{\rm visible}} 
 &=& \int_{0}^{\pi} dy\int d^4 \theta e^{-(T_0+T_0^\dag) \sigma} 
 \left[ Q^\dag_i e^{-V} Q_i 
 + c_{ij}\frac{H_0^\dag H_0 Q_i^\dag Q_j}{M_5^3} 
 \right] \delta(y-\pi) . 
\eea
Here we have assumed minimal K\"ahler potentials 
for the first term in each brackets for simplicity, 
$F_\phi=0$ is taken as an approximation suitable for 
the following discussion, 
$V$ is a vector superfield in the visible sector, 
$c_{ij}$ are flavor-dependent constants, and 
$H_0$ is the massless mode of $H$ given by 
\bea
H_0(x, y) = \frac{1}{N_0}e^{(\frac{3}{2} - c) T_0 k |y|} h_0(x), 
\eea 
with the normalization constant 
\bea 
|N_0|^2 = \frac{e^{(\frac{1}{2}-c)(T_0+T_0^\dag)k \pi}-1}{(1-2c) k}. 
\eea 
We take into account contributions only from massless mode, 
since contributions from massive modes are expected 
 to be exponentially suppressed by Yukawa potential. 
In terms of canonically normalized $H_0, Z$ and $Q_i$, 
the contact interactions in (\ref{hkahler}) and (\ref{vkahler}) 
are rewritten as  
\bea
{\cal L}_{{\rm hidden}} 
&\supset& \frac{1}{|N_0|^2} \int d^4 \theta 
 \frac{h^\dag_0 h_0 Z^\dag Z}{M_5^3}  , \\
{\cal L}_{{\rm visible}} 
&\supset& c_{ij} \frac{e^{(\frac{3}{2}-c)(T_0+T_0^\dag) k \pi} }
{|N_0|^2} \int d^4 \theta 
 \frac{h^\dag_0 h_0 Q_i^\dag Q_j}{M_5^3 } . 
\eea
The scalar masses induced by 1-loop corrections 
through the bulk hypermultiplet 
are roughly estimated as 
\bea
\Delta \tilde{m}^2_{ij} &\sim& 
\frac{1}{16 \pi^2} c_{ij}
\frac{e^{(\frac{3}{2}-c)(T_0+T_0^\dag) k \pi}}{|N_0|^4 M_5^6}
 \left| F_Z \right|^2 
\times V_{\rm{eff}}^{-2}  \nonumber  \\
&\sim& \frac{1}{16 \pi^2} c_{ij} m_{3/2}^2 \left( \frac{k}{M_4} \right)^2
\left( \frac{1 - 2c}
{e^{ (\frac{1}{2}-c) (T_0+T_0^\dag) k \pi}-1} \right)^2 
e^{(\frac{3}{2}-c)(T_0+T_0^\dag) k \pi}. 
\eea
Here we have used the relation $M_4^2 \sim M_5^3/k$ 
between the 4D Planck and the 5D Planck scales, 
$V_{\rm{eff}} \sim 1/k $ is the effective volume 
of the fifth dimension in the warped background metric 
by which the loop integral is expected to be cutoff physically, 
$1/16 \pi^2$ is a 1-loop suppression factor. 
For $c > \frac{3}{2}$ or $c < -\half$, 
$ \Delta \tilde{m}^2_{ij}$ is strongly suppressed. 
This is because, in the case with $c > \frac{3}{2}$, 
$H_0$ is localized around the hidden brane 
and the overlapping with the visible brane 
is exponentially suppressed. 
On the other hand, a zero mode $H_0$ is localized around 
the visible brane and the overlapping with the hidden brane 
is exponentially suppressed for $c<-\half$. 
Even for $c \sim 3/2, -1/2$, 
the contribution can be suppressed if $k \ll M_4$ 
or equivalently $k \ll M_5$ through the relation 
between Planck scales, $M_4^2 \sim M_5^3/k$. 
This is a natural situation when the bulk gravity is weak enough 
to be consistent with the classical treatment. 
For example, if we take $k \sim 0.1 M_5$, then $M_4^2 \simeq 10^3 k^2$, 
we find $\Delta \tilde{m}^2_{ij}/\tilde{m}^2_{{\rm AMSB}} 
\sim 10^{-3} \ll 1$ being consistent 
with current experimental results, 
where $\tilde{m}^2_{{\rm AMSB}} \sim (1/16 \pi)^2 m_{3/2}^2$  
is the anomaly mediation contribution. 

The gravity multiplet always exists in the bulk. 
Let us consider scalar masses induced 
through the bulk gravity multiplet loop corrections. 
This contribution is expected to be flavor blind 
since the fundamental interactions between fields on the branes 
and the bulk gravity multiplet are controlled by the SUGRA symmetry. 
In the flat background case, 
this contribution is directly calculated in \cite{GR} 
and the result is given by
\footnote{Another interesting contribution 
induced by corrections through the bulk gravity multiplet loop 
have been calculated in \cite{BGGLLNP,RSS}. 
This is found to be proportional to $1/(M_4 r_0)^3$ 
and sub-dominant. }
\bea
\label{5Dflat}
\Delta m^2_{{\rm 5D~flat}} \sim 
-\frac{1}{16 \pi^2} m_{3/2}^2 \frac{1}{(M_4 r_0)^2}. 
\eea
Unfortunately, this is the negative contribution 
and should be suppressed compared with anomaly mediation 
contributions to avoid tachyonic scalar fields. 
Although the corrections through the gravity multiplet
in the warped case has not yet explicitly calculated, 
we guess the result from the analogy in the flat case. 
As discussed in \cite{GR}, 
 the result of (\ref{5Dflat}) can be obtained 
 from the result of gravitino loop corrections in 4D SUGRA. 
It is known that in 4D SUGRA 
the scalar mass squared induced by gravitino 
1-loop corrections diverges quadratically. 
The result is given by 
\bea
\Delta m^2_{{\rm 4D~flat}} \sim \frac{1}{16 \pi^2} 
m_{3/2}^2 \frac{\Lambda^2}{M_4^2},    
\eea
where $\Lambda$ is a cutoff scale. 
The above result in 5D SUGRA case 
can be obtained by replacing the cutoff scale $\Lambda$ 
with the inverse of the extra-dimensional volume $1/r_0$. 
Recalling the Planck scale matching relations   
\bea
M_4^2 &=& M_5^3 r_0~({\rm flat~case}), \\
M_4^2 &\sim& \frac{M_5^3}{k}~({\rm warped~case}),
\eea
we naively expect that the scalar mass squared in 5D warped case 
is obtained by replacing $1/r_0$  with $k$ such as
\bea
\label{5dwarpscalar}
\Delta \tilde{m}^2_{{\rm 5D~warped}} 
\sim -\frac{1}{16 \pi^2} m_{3/2}^2 \left(\frac{k}{M_4} \right)^2 . 
\eea
As mentioned before, 
this negative contribution should be smaller than 
the anomaly mediation contributions. 
For example, if we take $k \sim 0.1 M_5$, then 
$M_4^2 \simeq 10^3 k^2$, 
we find $\Delta \tilde{m}^2_{{\rm 5D~warped}}/\tilde{m}^2_{{\rm AMSB}} 
\sim 10^{-3} \ll 1$. 

\section{Summary}
We have proposed a simple model of extra-dimensional radius stabilization 
in the supersymmetric Randall-Sundrum model. 
With only a bulk hypermultiplet 
and the source terms on each boundary branes, 
the radius stabilization has been succeeded 
through the SUSY vacuum conditions. 
The radion mass is found to be large enough, 
this radius stabilization is ensured even 
if the SUSY breaking effects are taken into account.  
Our model gives a remarkable advantage 
for model building in the SUSY brane world, 
since the radius can be stabilized independently 
of the SUSY breaking and its mediation mechanism. 
Our model may be applicable to many models. 
We find a reasonable parameter region where 
unwanted contributions to scalar mass squared 
through the bulk multiplets are suppressed enough. 

As a bonus of our model, 
if the bulk hypermultiplet $H$ is identified 
with the right-handed neutrino, and has couplings 
among Higgs doublet and the left-handed lepton doublet 
on the visible brane at $y=\pi$, 
we can naturally obtain a tiny neutrino mass 
through the mechanism proposed by Grossman and Neubert \cite{GN} 
with the hypermultiplet mass $c > \frac{3}{2}$.
In order to obtain realistic neutrino mass matrix, 
at least one extra hypermultiplet have to be introduced \cite{GN}. 
Such an extension is straightforward.

Finally, we comment on the dynamical origin of 
the source terms $J_{0,\pi}$. 
The source terms on each branes has the same form as 
Polonyi model \cite{Polonyi}. 
By introducing some strong coupling gauge theories 
with some superfields on each branes, 
we can easily construct a model 
where the source terms on each brane are dynamically generated 
through the strong gauge dynamics 
by the same manner as in \cite{IY,IT}. 
Here we give a rough picture of such models. 
Introduce a SUSY $SU(2)$ gauge theory with four doublets ($V_i$) 
on a brane ($y=0$ or $\pi$), and consider a superpotential 
\bea
W = \left. \frac{1}{\sqrt{M_5}}[V_i V_j] H \right|_{y=0,\pi} 
\eea
At low energies, the meson composite superfield $[V_i V_j] $ 
develops a non-zero VEV, $\langle [V_i V_j] \rangle = \Lambda^2$, 
through the quantum moduli deformation \cite{Seiberg}, 
where $\Lambda $ is the dynamical scale of 
the $SU(2)$ gauge theory. 
As a result, we obtain $J_{0, \pi} \sim \Lambda^2/{\sqrt{M_5}}$.

\vspace{1cm}

\begin{center}
{\bf Acknowledgments}
\end{center}
N.M. is supported 
by Special Postdoctoral Researchers Program at RIKEN. 
N.O would like to thank the particle theory group 
of National Center for Theoretical Sciences (NCTS) in Taiwan 
for warm hospitality during his visit to NCTS 
where some parts of this work were completed. 

\vspace{1cm}

\end{document}